\documentclass[12pt]{article}

\begin{document}

\begin{flushright}
YITP-SB-04-67
\end{flushright}

\bigskip

\begin{center}

{\Large \bf Energy Flow Observables\footnote{Based on a talk presented at the 44th INFN Workshop, ``QCD at  Cosmic Energies", Aug.29 - Sept.5, 2004, Erice, Italy.}}

\bigskip

{\large George Sterman}

\medskip

{\it C.N.\ Yang Institute for Theoretical Physics\\
Stony Brook University, SUNY\\
Stony Brook, New York 11794 -- 3840, U.S.A.}

\end{center}

\begin{abstract}
A few comments are made on the role of nonperturbative
and perturbative power corrections.  This is followed by a
description of energy flow observables and
correlations that may provide a
flexible approach to rapidity distributions in
hadronic scattering.
\end{abstract}

\section{From Factorized Cross Sections to Classical Fields}

This short talk describes a modest attempt to synthesize one or two of 
the ideas exchanged at this stimulating workshop.   I will try
to connect two themes relevant to the very highest energy collisions,
the saturation scale and rapidity distributions, with
perturbative QCD for hard-scattering cross sections.

The very high energies observed in cosmic ray collisions make it
natural to think of perturbative QCD,
but of course momentum transfers are by no means asymptotically
large in most inelastic collisions in the  upper atmosphere 
\cite{revie}, or at accelerators for that matter.
At extemely high energies, however, the same Lorentz
contractions of the target and projectile that
are at the basis of factorization may also result in very high
effective field strengths, which can provide a new dynamical energy,
referred to as the saturation scale \cite{satscale}.  The saturation 
scale acts in a sense
as a  mean momentum transfer for all partons in the dense medium.
For high enough densities and energies,
it can  be well into the perturbative region.  We can ask how
such a  phenomenon can arise in the language of perturbative
QCD for hard scattering.

In perturbative QCD,
the cross section for $A+B \rightarrow E({\it Q})+X$, with $A$ and $B$ 
hadrons
and $E(Q)$ an object in the final state of mass $Q$ (heavy particle, jet pair, 
etc.) can be
factorized.  In terms of parton distributions $f_{a/A}$ and a hard scattering 
function
$H_{ab\rightarrow E}$,  we have schematically \cite{cssrv}
\begin{eqnarray}
\sigma_{AB\rightarrow E}
=
\sum_{ab} f_{a/A}\, \otimes H_{ab\rightarrow E}\, \otimes\, f_{b/B}\, ,
\label{pmfact}
\end{eqnarray}
up to corrections that are suppressed
by a relative power of $1/Q$.
The convolution represented by $\otimes$ is an integral in
momentum fraction ($x$), and possibly also transverse momentum
\cite{ktfact}.
For ``minimum bias" events, or events
with one or more moderate momentum transfers
between partons, the power corrections
are not highly suppressed no matter how high the energy.
It is therefore worthwhile to review the origins of power
corrections as they appear in perturbative QCD.

An essentially exhaustive list of  power corrections derived from
perturbation theory for unpolarized scattering is:

\begin{itemize}

\item Strong coupling (renormalon) and/or vacuum corrections
can appear in the perturbative hard-scattering function through
nonconvergent behavior at high orders: $\alpha_s^n(Q)\, b_0^n\, n!$,
with $b_0$ the first coefficient of the  QCD beta function.  These
corrections often begin at $1/Q$ in semi-inclusive cross
sections \cite{irrrev}, but at $1/Q^2$ in single-particle
inclusive cross sections \cite{sv04}.

\item Mulitparton corrections, including parton transverse  momentum
effects involve two rather than one partonic degree of freedom
from one or both of the hadrons that participate in the hard scattering
\cite{htfact}.
These corrections may be put in a form analogous to (\ref{pmfact}),
and typically begin at $1/Q^2$ in unpolarized cross sections,
\begin{eqnarray}
\frac{1}{Q^2}\; \sum_{aa',b} F_{aa'/A}\, \otimes\, H_{aa',b\rightarrow 
E}\, \otimes\, f_{b/B}\, ,
\label{multipm}
\end{eqnarray}
where $F_{aa'/A}$ is a double distribution for degrees of 
freedom $a$ and $a'$, while $f_{b/B}$ is a standard single-parton
distribution.

\item Multiple scattering processes, with distinguishable components
to  the final state: $E=E_1+E_2$, can result from  independent 
hard scatterings
of different partons from both incoming hadrons \cite{multscat}.  These 
again
begin at $1/Q^2$,
\begin{eqnarray}
\frac{1}{Q^2}\; \sum_{aa',b} \bar{F}_{aa'/A}\, \otimes\, h_{ab\rightarrow
E_1}\, 
\otimes\, h_{a'b'\rightarrow E_2}
\otimes\, \bar{F}_{bb'/B}\, .
\label{multiscatt}
\end{eqnarray}
These are the corrections that correspond to multiple parton 
interactions,
as incorporated into models based on multiple minijet production.
The $\bar F$'s in (\ref{multiscatt}) are generally not the same as the
$F$'s in (\ref{multipm}).

\item Certain initial-state interactions beginning at $1/Q^4$ cannot
be written as the product of separate parton distributions for the
two hadrons at all \cite{qs}.  Such corrections still enjoy a 
factorization of the
hard scattering, which is initiated by two partons collinear
to the incoming hadrons, at leading relative power,
\begin{eqnarray}
\frac{1}{Q^4}\; \sum_{ab} {\mathcal F}_{ab/AB} \otimes \bar{H}_{ab\rightarrow E}\, .
\end{eqnarray}
These corrections, which do not allow a mutual factorization
in terms of universal parton distributions
in individual hadrons, are associated with
nonperturbative soft gluons.  Such soft gluons can couple to hard 
partons
originating from the colliding hadrons via
the QCD field strength.  In quantum perturbation theory \cite{brs},
as in solutions of the classical equations of  motion \cite{mv}, the
Lorentz contraction properties of the field strength play
an essential role.
  It is at this level that a classical color field
enters the perturbative factorization-based picture, as a
necessary completion of it.

\end{itemize}

Each of the extensions of leading-power factorized cross sections
appears in perturbation theory, in general requiring new
sets of nonperturbative degrees of freedom, 
 and the introduction of new distributions for 
multiple partons.
It therefore makes sense to ask whether we can
develop phonomenological tools to connect 
perturbative and nonperturbative dynamics in a controllable
and continuous fashion.
We may seek observables with adjustable parameters,
such that for some range of values perturbation theory is
accurate, while as we vary these parameters
outside the perturbative range, failures of perturbation theory may
provide insight into the relevant dynamics.
In the second section, I will suggest a class of observables
that are sensitive to energy flow in hadronic collisions 
to illustrate this approach.

\section{Energy Flow in Hadronic Collisions}

Plans for forward coverage 
at the LHC \cite{totem,castor}, and  the need to test shower event
generators \cite{revie} at cosmic energies both suggest the
usefulness of observables that are sensitive to
the global structure of particle production in
hadronic collisions.  Such observables are familiar
from $\rm e^+e^-$ annihilation as event shapes \cite{event}.
The classic set of event shapes includes the thrust, heavy jet mass,
and others.  Modified versions of these shapes
can be adapted to high-$p_T$ jets in hadronic
collisions \cite{eshadron}.  Our goal here is to point out
how event shapes can interpolate
between high $p_T$ and the forward direction.

It is generally not possible to take event shapes
over unchanged from leptonic annihilation to
the forward jets in
hadronic scattering cross sections.  This is
because initial-state radiation and forward
parton-parton scattering are highly singular in
the forward directions, due to the $1/\sin^4(\theta/2)$
behavior of the `Rutherford' cross section associated
 with the  exchange of gluons.

Nevertheless, it is possible that event
shapes with adjustable parameters may be
adapted to hadronic scattering, so that for
some range of values they can be written in
factorized form, similar to Eq.\ (\ref{pmfact}) above.   One such set of
event shapes with an adjustable parameter
is the set of ``angularities" \cite{angul},
\begin{eqnarray}
\tau_a(n) &=& \frac{1}{\sqrt{s}}\, \sum_{i\in n} \omega_i\,
\sin^a\theta_i\; \left(1-|\cos\theta_i|\right)^{1-a}
\nonumber\\
&=& \frac{1}{\sqrt{s}}\, \sum_{i\in N}\ k_{i\perp}\, {\rm e}^{-(1-a)|\eta_i|}\, .
\end{eqnarray}
For the case at hand, the angle $\theta$ and the rapidity 
$\eta$ are defined
relative to  the collision axis in the center of mass frame.
Defined this way, $0<\tau_a<1$.
We can construct a set of dimensionless global
energy flow observables for the collisions of
hadrons $A$ and $B$ in terms of angularities,
\begin{eqnarray}
M_{\mathrm AB}(N,\zeta)
=
S^{1-N/2}\; \sum_{{\mathrm states}\ n} \sigma_{\mathrm AB}(n)\, 
e^N_\zeta(n)\, ,
\label{mabdef}
\end{eqnarray}
where for each final state $n$ we define weighted transverse energies as
\begin{eqnarray}
e_\zeta(n) = \sum_{i\in n} k_{i\perp}{\rm e}^{-\zeta|\eta_i|}\, ,
\label
{edef}
\end{eqnarray}
with parameter $\zeta\equiv 1-a$.

The angularities, and hence the energy flow variables
$e_\zeta$ are defined so that as $\zeta$ increases or decreases,
the contributions of particles in the forward directions 
$|\eta|\rightarrow\infty$
are weighted less or more.   To show how  this works, consider
the lowest order contribution to $M_{\mathrm AB}$ from a single parton-parton
scattering, with two high-$p_T$ particles (or jets) in
the final state, one of which is at fixed rapidity $\eta$
in the hadronic center of mass.  If the momentum transfer, and therefore
transverse momentum is large ($\eta$ is fixed), this contribution
is simply the value of $e_\zeta^N$ times the lowest order
jet cross section, where $e_\zeta$ gets contributions
from both final-state particles.   
The cross section depends in the usual way
on parton distributions and hard scattering functions.
For simplicity and to study the most singular
behavior,  we show only the lowest-order $t$-channel exchange 
contribution,
\begin{eqnarray}
|A_{ij}(\hat s,\hat t)|^2 \rightarrow C_{ij}(\alpha_s/\pi)^2\, 
\left({\hat s\over \hat t}\right)^2
=
C_{ij}(\alpha_s/\pi)^2\, {\rm e}^{2\eta^*}\cosh^2\eta^*\, ,
\end{eqnarray}
with $C_{ij}$ a constant,
for parton flavors $i$ and $j$, and with $\eta^*$ the
rapidity of either outgoing parton in the partonic center of mass.  

We can write the full lowest-order contribution to
$M_{\mathrm AB}$ as
an integral over hadronic c.m.\ rapidity $\eta$ and
$x_T\equiv 2p_T/\sqrt{S}$,
\begin{eqnarray}
M_{\mathrm AB}^{\mathrm (t\ channel)}(N,\zeta)
&=&
\sum_{ij}C_{ij}
  \int_{-\infty}^{\infty} d\eta\, 
\int_0^{{\rm e}^{-|\eta|}}dx_T\, x_T^{-1}
\nonumber\\
&\ & \hspace{-30mm} \times
\left({\alpha_s\left({x_T\sqrt{S}\over 2}\right)\over\pi}\right)^2
  \int_{-{1\over 2}\ln\left( {2{\rm e}^{-\eta}\over 
x_T}-1\right)}^{{1\over 2}\ln\left( {2{\rm e}^{\eta}\over 
x_T}-1\right)} d\eta^*\; {\rm e}^{2\eta^*}\; \left[\, x_T{\rm e}^{-\zeta|\eta|}\, +\, x_T
{\rm e}^{-\zeta|\eta-2\eta^*|}\, \right]^N \nonumber
\\
&\ & \hspace{-30mm}\times
\ f_{i/A} \left(x_T{\rm e}^{\eta-\eta^*}\cosh\eta^*,{x_T\sqrt{S}\over 
2} \right)\;
f_{j/B} \left(x_T{\rm e}^{-\eta+\eta^*}\cosh\eta^*,{x_T\sqrt{S}\over 
2}\right)\, . \nonumber\\
\label{lomab}
\end{eqnarray}
This serves as a perturbative description of energy flow.  
The two rapidity-dependent terms in the square brackets give
the contributions of the two final-state particles, with equal $x_T$, to
the weight at lowest order.
For small $N$ there is a  
manifest
singularity at $x_T=0$, which is strengthened by the 
exponential growth of the $t$-channel amplitude
 in $\eta^*$.  Choosing as above the 
factorization
scale as $\mu_F=p_T=x_T/2\sqrt{S}$, the precise perturbative predictions
will also depend on the low-$x$ behavior of the parton distributions at
low $\mu_F$, but for $N$ and $\zeta$ large enough, $M_{\mathrm AB}$  is
perturbatively calculable, and insensitive to this
behavior up to power corrections.   
(This is similar to the situation for transverse momentum
distributions of Drell-Yan pairs.)
Such moments of the energy flow distribution are calculable for any
fixed rapdity $\eta$.  

The observables described above are not limited to
perturbative calculation based on collinear factorization.  They
can in principle be used to test 
predictions of any model of high energy scattering that
provides detailed predictions for energy flow in
the final state.  
By varying the two parameters, $N$ and
$\zeta$, and exploiting the forward coverage 
planned for the LHC, it may be possible, for example,  to quantify 
correlations of activity in the forward region with hard scattering
\cite{fra04}.  
Other examples may include predictions based on
$k_T$-factorized cross sections, as have been advocated especially
for nuclear collisions, in addition to those of event
generator models.   Angularity-based analyses of the global properties
of final states in hadron-hadron collisions may be useful, but
they are only a first proposal.
The main message of this talk is that the large rapidity
coverage plannned for the LHC will open the way to varied studies
  and invite the development of new analyses that probe
  the formation of QCD final states.
  
  \subsection*{Acknowledgments}

I would like to thank the organizers of the Workshop for
inviting me, and
the Ettore Majorana Institute for support.  This work is supported 
in part by the National Science Foundation, grants PHY-0098527 
and PHY-0354776.

\end{document}